\documentclass[preprint,showpacs,amsmath,amssymb,aps,11pt]{revtex4}

\usepackage{graphicx}
\usepackage{dcolumn}
\usepackage{bm}

\begin{document}


\title{On Hawking Radiation as Tunneling Phenomenon with Dissipation Effect}

\author{Haryanto M. Siahaan}
 \affiliation{St. Aloysius School,\\ Jalan Batununggal Indah II/30, Bandung 40266,\\ West Java, Indonesia} 
 \email{anto_102@students.itb.ac.id}


\begin{abstract}
In this letter, we derive the Hawking temperature in the tunneling method by considering the system is open to the environment. This consideration yields the temperature which is higher than the standard one. The correction of temperatures depends on spectral density of bath oscillators which is responsible for dissipation effect in our analysis.
\end{abstract}

\maketitle

Hawking radiation demonstrates connection among gravity, quantum theory, and thermodynamics interpretation of black hole mechanics \cite{1}-\cite{3}. Semi-classical method in the quantum tunneling picture in deriving 
Hawking temperature has been introduced by Volovik in \cite{Volovik1999a} and also \cite{Volovik1999b}. He used the analogy between superfluidity and Hawking radiation. Earlier of this method  was applied to the Unruh effect (and its condensed matter analogs) \cite{Volovik1992}. It was also applied to Zeldovich-Starobinskii effect (quantum tunneling from the ergosurface of  rotating black hole) and to its condensed matter analogs in \cite{CalogeracosVolovik1999}. The details can be found in \cite{Volovik2008}.\par
The semiclassical tunneling approaches have been developed for the astronomical black holes by Parikh--Wilczek and Padmanabhan et. al., i.e. null geodesic \cite{4} and complex paths \cite{5}--\cite{19} methods. The methods have been being active research topics in the past decade, e.g. \cite{6}-\cite{10}, and even to cosmology, e.g. \cite{Cai}. Recently, the authors of \cite{16} have unified the tunneling formalism and anomaly method through chirality and used the tunneling formalism to derive the black body spectrum at \cite{17}. The main idea of these methods is outgoing probability (emission rate) calculation of a particle from inside of black hole horizon to infinity along classically forbidden trajectory. Then one can equalize it with the Boltzman factor to get the corresponding Hawking temperature.
\par
Since we are discussing about tunneling process where the energy of outgoing particle is less than the potential that will be tunneled, the corresponding action of particle will become complex. Emission rate of this outgoing particle is given by
\begin{eqnarray}
	\Gamma  \sim \exp \left( { - 2{\mathop{\rm Im}\nolimits} S} \right)\label{eq:1}
\end{eqnarray}
Understanding tunneled particle as the phenomenon of quantum systems open to environment can be formidable challenge. The particle, for instance, can exchange energy with the surrounding medium. This leads to dissipation effect to the tunneling process \cite{11}-\cite{12}. It is clear that the consideration of dissipation effect for tunneled particle in Hawking radiation would give a correction to the corresponding Hawking temperature. This idea will be worked out via null geodesic method \cite{4}. We use the Caldeira-Leggett (CL) model \cite{12} to describe open quantum system where the Hamiltonian contains system, bath, and coupling between them (system-bath).
\par
We start by writing the Hamiltonian in CL model
\begin{eqnarray}
	H = H_S  + H_B  + H_{S - B} \label{eq:2}
\end{eqnarray}
where\footnote{S and B denote system and bath respectively}
\begin{eqnarray}
	H_S  = \frac{{p^2 }}{{2m}} + V\left( r \right)\label{eq:3},
\end{eqnarray}
\begin{eqnarray}
	H_B  = \sum\limits_{n = 1}^N {\left( {\frac{{p_n ^2 }}{{2m_n }} + \frac{{m_n }}{2}\omega _n ^2 x_n ^2 } \right)}\label{eq:4},
\end{eqnarray}
and
\begin{eqnarray}
	H_{S - B}  =  - r\sum\limits_{n = 1}^N {c_n x_n }  + r^2 \sum\limits_{n = 1}^N {\frac{{c_n ^2 }}{{2m_n \omega _n ^2 }}}\label{eq:5}. 
\end{eqnarray}
$\left( {r,p} \right)
$ and $\left( {x_n ,p_n } \right)
$ are both pairs of coordinate and its conjugate momentum respectively, i.e. $p = m\dot r
$ and $p_n  = m_n \dot x_n 
$. We have used the notation $\dot a \equiv {{da} \mathord{\left/
 {\vphantom {{da} {dt}}} \right.
 \kern-\nulldelimiterspace} {dt}}
$ and variables $r$, $x_n$, $m$, $\omega_n$, and $m_n$ denote the radial coordinate of tunneled particle, coordinate of bath harmonic oscillators (will be shortened by b.h.o. from now on), mass of particle, frequency and mass of b.h.o. respectively. Without loss of generality, one also can choose \cite{12} the coupling constant between system and bath $c_n$ as the following
\begin{eqnarray}
	c_n  = m_n \omega _n ^2\label{eq:6}.
\end{eqnarray}
\par
In the Hamiltonian (\ref{eq:2}), we use directly the momentum $p$ with subscript $r$ to denote that we are working in radial coordinate that is suitable with our later analysis on $2$-dimensional black hole. The system without coupling may have certain equilibrium positions which usually will be shifted if one couples it to other degrees of freedom. In this letter, we consider that both coordinates $r$ and $x_n$ are in equilibrium. It is possible by setting
\begin{eqnarray}
	\frac{{\partial H}}{{\partial x_n }} = 0.\label{eq:7}
\end{eqnarray}
This condition ensures that the minimum of Hamiltonian would be determined only by the bare potential $V\left( r \right)
$, i.e. ${{\partial H} \mathord{\left/
 {\vphantom {{\partial H} {\partial r}}} \right.
 \kern-\nulldelimiterspace} {\partial r}} = {{\partial V\left( r \right)} \mathord{\left/
 {\vphantom {{\partial V\left( r \right)} {\partial r}}} \right.
 \kern-\nulldelimiterspace} {\partial r}}
$.
Condition (\ref{eq:6}) yields a relation between system and bath coordinate which can be read as
\begin{eqnarray}
	x_n  = \frac{{c_n }}{{m_n \omega _n ^2 }}r.\label{eq:8}
\end{eqnarray}
The last equation therefore simply means that in equilibrium, system and bath coordinates fulfill a certain relation\footnote{I thank G. -L. Ingold for his explanation on this subject.}. Furthermore, relation (\ref{eq:7}) would lead us to an expression
\begin{eqnarray}
	\frac{{dp_n }}{{dp}} = \frac{{c_n }}{{m\omega _n ^2 }},\label{eq:9}
\end{eqnarray}
which will be useful in our later calculation.
\par
In the null radial geodesic method, we get the dynamic of particle $\dot r
$ from the null geodesic condition $ds^2  = d\Omega _2 ^2  = 0
$ for general metric
\begin{eqnarray}
	ds^2  =  - f\left( r \right)dt^2  + g\left( r \right)^{ - 1} dr^2  + r^2 d\Omega _2 ^2\label{eq:10} 
\end{eqnarray}
with time like Killing vector. $d\Omega _2 ^2 
$ denotes the metric for $2$-sphere. To remove the singularity at the horizon, $r = r_h 
$ i.e. $f\left( {r_h } \right) = g\left( {r_h } \right) = 0
$, one can perform the transformation (see \cite{vagenas})
\begin{eqnarray}
	dt \to dt - \sqrt {\frac{{1 - g\left( r \right)}}{{f\left( r \right)g\left( r \right)}}} dr.\label{eq:11}
\end{eqnarray}
By this transformation, one get the metric that can be read as
\begin{eqnarray}
	ds^2  =  - f\left( r \right)dt^2  + 2f\left( r \right)\sqrt {\frac{{1 - g\left( r \right)}}{{f\left( r \right)g\left( r \right)}}} dtdr + dr^2  + r^2 d\Omega _2 ^2\label{eq:12} 
\end{eqnarray}
which is known as Painleve coordinates. The approximation of $\dot r
$ can be obtained by expanding $f\left( r \right)
$ and $g\left( r \right)
$ near horizon, $r_h 
$, and the result can be written as the following
\begin{eqnarray}
	\dot r \simeq \frac{1}{2}\sqrt {f'\left( {r_h } \right)g'\left( {r_h } \right)} \left( {r - r_h } \right).\label{eq:13}
\end{eqnarray}
We have used the notation $a' \equiv {{da} \mathord{\left/
 {\vphantom {{da} {dr}}} \right.
 \kern-\nulldelimiterspace} {dr}}
$ in (\ref{eq:13}). The last expression tells us that the dynamic of outgoing particle from inside of horizon to infinity is governed by the geometry of the black hole.
\par
Now, we focus on total Hamiltonian as expressed in (\ref{eq:3}). One can write
\begin{eqnarray}
	\frac{{dH}}{{dp}} = \frac{{\partial H}}{{\partial p}} + \frac{{\partial H}}{{\partial p_n }}\frac{{dp_n }}{{dp}}.\label{eq:14}
\end{eqnarray}
By combining (\ref{eq:14}) with the Hamilton equation $\dot r = {{dH} \mathord{\left/
 {\vphantom {{dH} {dp}}} \right.
 \kern-\nulldelimiterspace} {dp}}
$ (evaluated at $r$), thus we get the relation
\begin{eqnarray}
	\frac{{dH}}{{dp}} = \dot r\left( {1 + \sum\limits_{n = 1}^N {\frac{{c_n ^2 }}{{m_n m\omega _n ^2 }}} } \right).\label{eq:15}
\end{eqnarray}
Since $c_n 
$, $m_n$, and $\omega_n$ are microscopic variables, we can present them in spectral density \cite{12} that is written as
\begin{eqnarray}
	J\left( \omega  \right) = \pi \sum\limits_{n = 1}^N {\frac{{c_n ^2 }}{{2m_n \omega _n }}\delta \left( {\omega  - \omega _n } \right)}.\label{eq:16} 
\end{eqnarray}
By the use of the choice in (\ref{eq:6}) that yields
\begin{eqnarray}
	J\left( \omega  \right) = \frac{\pi }{2}\sum\limits_{n = 1}^N {m_n \omega _n ^3\delta \left( {\omega  - \omega _n } \right)}, 
\end{eqnarray}
we can write (\ref{eq:15}) in a more compact form
\begin{eqnarray}
	\frac{{dH}}{{dp}} = \dot r\left( {1 + \frac{C}{m}} \right),\label{eq:17}
\end{eqnarray}
where
\[
C \equiv \sum\limits_{n = 1}^N {m_n }  = \frac{2}{\pi }\int\limits_0^\infty  {\frac{{J\left( \omega  \right)}}{{\omega ^3 }}d\omega }. 
\]
The mass of particle that is written as $m$, in (\ref{eq:17}) could be replaced by $\varepsilon 
$ (energy of outgoing particle) since we are working in natural dimensions where Newton's constant, velocity of light in the vacuum, and Boltzman constant are unity, i.e. $G = c = k_B  = 1
$.
\par
$C$ is a constant that represents the total mass of b.h.o. which is physically assumed to be finite. Also, $C$ can be viewed as the result of interaction between system and bath in our analysis. This term represents the dissipation effect in the quantum tunneling process as the basic mechanism for Hawking radiation in this letter. It depends on spectral density $J\left( \omega  \right)
$ which in dissipative quantum analysis plays a role as macroscopic variable. The most frequently used spectral density is the Ohmic one, $J\left( \omega  \right) = \eta \omega 
$ where $\eta 
$ is a constant. Indeed, using this type of spectral density directly into our total mass $C$ would lead to such infinity for low frequency $\omega$. Thus, one can apply the same method to avoid this singularity by introducing the cut off frequency $\omega_c$ in the integration, and finally yields a finite total mass as expected.
But we will not discussed any further on what kind spectral density that is suitable for Hawking radiation in the tunneling formalism. Our reason is there is no physical data that can be compared. Our aim in this letter is rather to show how to incorporate the dissipation effect in Hawking radiation analysis as open quantum system.
\par
To get the Hawking temperature in this analysis, we start by writing the imaginary action for outgoing particle
\begin{eqnarray}
{\mathop{\rm Im}\nolimits} S = {\mathop{\rm Im}\nolimits} \int\limits_{r_{in} }^{r_{out} } {pdr}  = {\mathop{\rm Im}\nolimits} \int\limits_{r_{in} }^{r_{out} } {\int\limits_0^p {dp'dr} }  = {\mathop{\rm Im}\nolimits} \int\limits_{r_{in} }^{r_{out} } {\int\limits_0^\varepsilon  {\frac{{d\varepsilon '}}{{\dot r\left( {1 + {C \mathord{\left/
 {\vphantom {C m}} \right.
 \kern-\nulldelimiterspace} m}} \right)}}dr} }.\label{eq:18}
\end{eqnarray}
We have replaced the infinitesimal Hamiltonian $dH$ in (\ref{eq:18}) with   since the change of energy of system is depend on the energy of outgoing particle. By combining the results in (\ref{eq:13}) and (\ref{eq:17}), we can write the imaginary action as has been shown in (\ref{eq:18}) as the following
\begin{eqnarray}
	{\mathop{\rm Im}\nolimits} S = {\mathop{\rm Im}\nolimits} \int\limits_{r_{in} }^{r_{out} } {\int\limits_0^\varepsilon  {\frac{{2\varepsilon 'd\varepsilon 'dr}}{{\left( {\varepsilon ' + C} \right)\sqrt {f'\left( {r_h } \right)g'\left( {r_h } \right)} \left( {r - r_h } \right)}}} }.\label{eq:19}
\end{eqnarray}
One can simply perform the $d\varepsilon'$ integration, and for $dr$ integration one can perform such contour integration for upper half complex plane to avoid the singularity at $r_h$. We consider $\sqrt {f'\left( {r_h } \right)g'\left( {r_h } \right)} 
$ that is generally as a function of black hole mass \footnote{For Schwarschild black hole, $r_h  = 2M
$.} $M$ would not depend on $\varepsilon '
$. The reason is when the tunneling is going on, the mass of the black hole is still unchanged into $M - \varepsilon '
$.\par
The result of integration in (\ref{eq:19}) can be read as
\begin{eqnarray}
	{\mathop{\rm Im}\nolimits} S = \frac{{2\pi \left( {\varepsilon  - C \ln \left( {\varepsilon  + C} \right)} \right)}}{{\sqrt {f'\left( {r_h } \right)g'\left( {r_h } \right)} }}.\label{eq:20}
\end{eqnarray}
Finally, by using the prescription proposed in \cite{4}, we can get the Hawking temperature after equating the emission rate as described in (\ref{eq:1}) with the Boltzman factor,
$\exp \left( { - {{\varepsilon \hbar } \mathord{\left/
 {\vphantom {{\varepsilon \hbar } T}} \right.
 \kern-\nulldelimiterspace} T}} \right)
$. The Hawking temperature as result from our analysis can be written as
\begin{eqnarray}
	T_H  = \frac{{\hbar \sqrt {f'\left( {r_h } \right)g'\left( {r_h } \right)} }}{{4\pi \left( {1 - \left( {{C \mathord{\left/
 {\vphantom {C \varepsilon }} \right.
 \kern-\nulldelimiterspace} \varepsilon }} \right)\ln \left( {\varepsilon  + C} \right)} \right)}}.\label{eq:21}
\end{eqnarray}
A modification of particle action for emission rate expression in equation (\ref{eq:19}) had been performed by several authors [e.g. \cite{6}, \cite{borun}, and \cite{doug}]. They used $
\oint {p_r dr} 
$ (which is canonically invariant) rather than $
\int {p_r dr} 
$. This choice yields the value of temperature to be twice the correct one. Indeed, our analysis can also be performed in this scheme. 
\par
Our final finding in (\ref{eq:21}) would indeed change into standard Hawking temperature \cite{6}-\cite{8} for $C=0$. The condition $C=0$ can be obtained by setting $m_n=0$ which means no dissipation source in the Hamiltonian (\ref{eq:2}), i.e. $H = H_S 
$. We find the correction 
\begin{eqnarray}
	\left( {1 - C\varepsilon ^{ - 1} \ln \left( {\varepsilon  + C} \right)} \right)^{ - 1} \label{eq:22}
\end{eqnarray}
for the standard Hawking temperature in the tunneling formalism $T_H  = \left( {4\pi } \right)^{ - 1} \hbar \sqrt {f'\left( {r_h } \right)g'\left( {r_h } \right)} 
$. To get a definite positive temperature, a physical condition that must be obeyed for $\varepsilon$ and $C$ can be written as the following
\begin{eqnarray}
	\ln \left( {\varepsilon  + C} \right) < \frac{\varepsilon }{C}.\label{eq:23}
\end{eqnarray}
This condition is physically reasonable. For example, if one consider the correction (\ref{eq:22}) would be a small valued one, thus $\ln \left( {\varepsilon  + C} \right) <  < C^{ - 1} \varepsilon 
$. This means the total mass of entire b.h.o must be much less than the energy of tunneled particle. This picture is physically acceptable.\par
Application of open quantum system analysis which rises the dissipation effect to the tunneling processes in condensed matter (such as superconductor, superfluidity, molecular transistors, etc.) has been performed by some authors \cite{condmat}. Volovik also had shown the close analogy between superfluidity and Hawking radiation \cite{Volovik1999a} \cite{Volovik1999b}. Most realistic black hole will not be in vacuum, but will have some matter 
around
such as from a companion star or interstellar medium. By these hints, we are convinced that (real) Hawking radiation will also exhibit the dissipation effect.\par

\par
Future investigation that would be interesting to be pursued further is finding the physical reason from the fluctuating metric point of view \cite{13} that is responsible to govern the dissipation effect in tunneling process. In ref. \cite{13}, York used physical metric fluctuation as the source of Hawking radiation. The very well known dissipation-fluctuation theorem \cite{14}-\cite{15} perhaps can be used to accommodate the idea. In this letter we just use the model by Caldeira-Leggett in explaining mechanism for transferring energy between tunneled particle with the surrounding environment. Another derivation (but in the same spirit with our analysis) via another semiclassical method for quantum tunneling Hawking radiation (Padmanabhan et. al. \cite{5}) also can be investigated. But perhaps it will need tedious calculation in deriving corresponding effective action for outgoing particle via path integral \cite{11}.\par 

\begin{center}
\textbf{Acknowledgements}
\end{center}
I would like to thank my mentor, Triyanta from THEPI division FMIPA-ITB, for valuable discussions and his supports. It is a pleasure to thank C. -L. Ingold, G. E. Volovik, T. Pilling, D. Singleton, S. Shankaranarayanan, and B. R. Majhi for their correspondences. I am grateful to Elias C. Vagenas for his comments and pointing out some errors in the previous version of this paper. My apology to G. E. Volovik for missing his works (\cite{Volovik1999a}--\cite{Volovik2008}) in the previous version of this paper.

\end{document}